\documentclass[traditabstract]{aa}
\usepackage[utf8]{inputenc}
\usepackage{amsmath}
\usepackage{amssymb}
\usepackage{microtype}
\usepackage{natbib}
\usepackage{graphicx}
\usepackage{txfonts}
\usepackage{color}
\usepackage{multirow}
\usepackage{xspace}
\bibpunct{(}{)}{;}{a}{}{,} 

\usepackage{xcolor}

\usepackage[colorlinks = true,
            linkcolor = blue,
            urlcolor  = blue,
            citecolor = blue,
            anchorcolor = blue]{hyperref}

\begin{document}

   \title{3D-$N_{\rm H}$-tool}

   \author{V. Doroshenko\inst{1}\thanks{E-mail: victor.doroshenko@gmail.com}}
\authorrunning{author}

   \institute{
   \inst{1}Universit{\"a}t T{\"u}bingen, Institut f{\"u}r Astronomie und Astrophysik T{\"u}bingen, Sand 1, 72076 T{\"u}bingen, Germany}
  \abstract{Absorption of light is one of the main selection effects limiting our ability to detect celestial sources, and ultimately, appearance of the sky across most of the electromagnetic spectrum. Recent advances in quantity and quality of available observational data and analysis methods led to major improvements in resolution, depth and fidelity of 3D dust distribution and extinction maps. The Galactic plane remains, however, essentially ``terra incognita'' beyond distances of a few kilo-parsecs due to the strong absorption in optical and near-infrared bands. Here I attempt to address this issue and present a 3D-$N_{\rm H}$-tool to estimate line of sight reddening and X-ray absorption column combining state of the art optical extinction and dust emission maps, and the results of dispersion measure modeling based on radio pulsar observations. The resulting maps are calibrated using independent absorption tracers and are accessible to general community via a convenient web-interface and full data cube.
  }
   \maketitle

%

\section{Introduction}
The importance of interstellar absorption for observational astronomy is hard to overstate. Extinction of starlight was one of the first selection effects recognized in astronomy and remains a major factor limiting our ability to study the vast majority of stars in our own Galaxy (approximately 99\%) for the foreseeable future \citep{2023arXiv230612894M}. 
Cosmic dust, responsible for the absorption in the optical, ultraviolet, and X-ray bands, also traces the densest parts of the interstellar medium (ISM). These regions are where star formation is most active. Thus, cosmic dust provides a unique perspective on the life cycles of stars, stellar feedback, chemical evolution, and the enrichment of the ISM with heavier elements, sheds light on planet formation and, ultimately, the origin of life. 

Historically, ISM studies have been among the most active fields in observational astronomy \citep{1980A&AS...42..251N, 1998A&A...340..543V}. The advent of modern large-scale spectroscopic and astrometric surveys, coupled with developments in analysis techniques, quality of stellar atmosphere models, and available computing resources, has led to significant breakthroughs in the field in recent years.
In particular, it became possible to construct high fidelity 3D optical extinction maps using multi band photometry and spectroscopy provided by wide area surveys \citep{2003A&A...409..205D, 2006A&A...453..635M,2014MNRAS.443.2907S, 2022MNRAS.517.5180Z}, most recently complemented by \textit{Gaia} astrometry  \citep{2018MNRAS.476.2556Q, 2019A&A...628A..94A, 2021A&A...649A...1G, 2023arXiv230801295E}. This allowed to push spatial resolution of 3D dust maps to parsec scales in Solar vicinity  (which, however, drops quickly beyond distances of $1-2$\,kpc due to limited sensitivity and accuracy of measured parallaxes (i.e. roughly limited to E23 volume). Use of deeper optical and near-infrared data allows to expand volume accessible to mapping to a few kilo-parsecs \citep{2019ApJ...887...93G,2022A&A...664A.174V}, but becomes increasingly challenging as large number of objects required for statistical inference of high-resolution extinction maps is simply not observable with current surveys. At this point other methods have to be employed, i.e. studies of cepheides, RR Lyrae, Miras/LPVs, red-clump stars, background galaxies and so on \citep[for review see][]{2023arXiv230612894M}. Still, our ability to probe distribution of stars and dust in the Galactic plane using optical and near infrared data remains extremely limited for distances beyond few kilo-parsecs with current optical instruments. 

Beyond this limit, direct detection of dust emission in the infrared and microwave bands \citep{1998ApJ...500..525S,2016A&A...596A.109P} can be used to study dust content of the ISM. However, only total dust quantity integrated along the line of sight is accessible. Interpretation of far-infrared observations also suffers from the necessity to disentangle Galactic and extra-Galactic components \citep{2016A&A...596A.109P,2023ApJ...958..118C}. Finally, the conversion of dust emission maps to optical reddening or X-ray absorption is not trivial and depends on certain assumptions requiring independent calibration. Nevertheless, far infrared reddening maps are commonly used as a tracer of total Galactic extinction \citep{2023arXiv230617749S} and, in fact, it is sometimes the only option. 

The distribution of gas fraction of the ISM along the line of sight can also be probed through radio observations. In particular, line spectroscopy \citep{1990ARA&A..28..215D, 2001ApJ...547..792D, 2016A&A...594A.116H} can be used to trace HII and CO distribution, provided that the Galactic rotation curve is known. The CO emission is especially interesting in this context as it traces the colder ISM region where most of the absorption occurs and thus can be contrasted with other tracers. However, uncertainties in the assumed rotation curve lead to substantial differences in results by different authors and for different tracer lines \citep{2008ApJ...677..283P,2021A&A...655A..64M,2023A&A...671A..54M}.

Finally, spatial distribution of the absorbing material in the Galaxy can be recovered from modeling of the observed dispersion measure (DM) of fast radio bursts (FRBs) and pulsars \citep{2002astro.ph..7156C,2017ApJ...835...29Y}. The advantage is that robust independent distance estimates are available for many radio sources, and thus no assumptions regarding the rotation curve are needed. However, the number of pulsars and FRBs is comparatively low. As a result, it is not possible to recover the density field directly. Instead, it has to be represented as a set of large-scale components such as the Galactic disk, spiral arm structure, and halo \citep{2002astro.ph..7156C,2017ApJ...835...29Y}. Nevertheless, DM-based 3D gas distribution maps are commonly (and successfully) used in radio astronomy as a distance proxy, for instance, for radio pulsars \citep{2005AJ....129.1993M}. 
An important caveat of radio measurements in the context of absorption at shorter wavelengths (both DM and line-based), is that they primarily trace the distribution of hot (DM and HII) or cold (CO) gas. However, the distribution of dust, which is mostly responsible for optical extinction and X-ray absorption, is known to be different. This discrepancy limits the accuracy of radio-based maps as tracers of optical reddening or X-ray absorption columns, unless additional modeling is carried out to account for variations in dust content and chemical composition across the Galaxy \citep{2013MNRAS.431..394W}.
Nevertheless, integrated HII distribution maps \citep{1990ARA&A..28..215D, 2016A&A...594A.116H} are commonly used in X-ray astronomy to estimate the Galactic absorption column density\footnote{via the HEASOFT NH-tool and \url{https://heasarc.gsfc.nasa.gov/cgi-bin/Tools/w3nh/w3nh.pl}}. The DM also appears to be correlated with the absorption column \citep{2013ApJ...768...64H}, suggesting that DM-based models could potentially be used to estimate the absorption column, although this approach is not commonly adopted. Nevertheless, optical and IR-based extinction maps are arguably better proxies for absorption in X-rays. However, their use in X-ray astronomy is limited by the lack of convenient tools to access absorption maps beyond those provided by the HII-based NH-tool from NASA's HEASARC$^{1}$.

Here I aim to address this shortcoming by providing a convenient 3D-$N_H$-tool providing an estimate of X-ray absorption column and optical reddening along the line of sight throughout the Galaxy. To obtain the estimate, use a combination of state of the art 3D optical extinction and dust emission maps, and results of the  dispersion measure (DM) modeling based on radio pulsar observations from the literature. The results are calibrated using several independent absorption tracers and can be used to estimate X-ray absorption column, visual and $K$-band extinctions and reddening.

The rest of the paper is organised as follows. In section~\ref{sec:input_maps}, data sources and their combination used to obtain the pan-Galactic 3D optical reddening and X-ray absorption maps are described. In section \ref{sec:calibration}, I compare the resulting maps with several independent absorption estimates to calibrate the derived 3D maps. Finally, in section \ref{sec:nhtool}, I describe implementation of the web tool and then summarise main results of this work in section~\ref{sec:summary}.
The tool itself accessible to general community via a convenient web-interface\footnote{\url{http://astro.uni-tuebingen.de/nh3d}}. Full data cube and code to obtain it are also available at \textit{Zenodo}\footnote{\url{https://zenodo.org/records/10779060}}.

\section{Absorption tracers and their combination}
\label{sec:input_maps}
As previously mentioned, Galactic absorption can be probed through several independent tracers. These include optical reddening, integrated far-infrared to radio dust and gas emission, and radio dispersion measure. The highest fidelity near the Sun is provided by optical reddening maps. Moreover, optical reddening maps already account for most of the absorbing material away from the Galactic plane. Indeed, the vertical distribution of the absorbing material is known to be confined to a few hundred parsecs from the Galactic plane, making it accessible to \textit{Gaia} \citep{2021A&A...649A...1G,2023arXiv230801295E} and other surveys. However, the situation closer to the Galactic plane is different where depth of current optical surveys is limited to a few kilo-parsecs, and thus also the reddening can not be recovered. The primary goal of the current work is to extend the available high-resolution 3D optical extinction (or reddening) maps in the Galactic plane region. This is essential
for population studies in X-ray and other bands, even if quality of the resulting maps can not be comparable to those in the Solar vicinity.
To achieve this goal I rely on far-infrared dust emission maps to estimate total absorption along the line of sight in the Galactic plane. Additionally, information from radio DM models is used to derive the distribution of the material along the line of sight beyond the volume covered by optical maps.
In this section, I provide a brief description of the data sets used and how they can be combined to assess dust distribution throughout the Galaxy.

\subsection{Optical and near-infrared extinction}
\label{sec:input_maps_optical}
The optical reddening map by \cite[E23 further on in the text]{2023arXiv230801295E} is the most recent and highest resolution 3D ISM map covering full sky. E23 apply information theory algorithms to derive distribution of absorbing material using \cite{2023MNRAS.524.1855Z} extinction estimates for over 200\,millions of objects based on optical photometry and medium-resolution spectroscopy by \textit{Gaia, unWISE} and \textit{2MASS} surveys and \textit{Gaia} astrometry. The map covers distances up to 1.25 kpc, with an extended version reaching up to 1.95 kpc. The map is stored as a set of \texttt{HEALPIX NSIDE=256} \citep{2005ApJ...622..759G} differential extinction maps for set of geometrically spaced distances from approximately 40 pc to 2 kpc. This corresponds to the constant angular resolution of 14$^\prime$ and resolution in radial direction from 0.4 pc to 12 pc depending on distance. The map reports monochromatic extinction in the units used by \cite{2023MNRAS.524.1855Z}, effectively making it a reddening map which can be converted to standard $E(B-V)$ units through multiplication by factor 2.8/3.1 (i.e. assuming the standard extinction law $R_V=3.1$). In the following sections, I use an integrated version of the map obtained with the \texttt{Edenhofer2023Query} function from the \textit{dustmaps} package \citep{2018JOSS....3..695M} converted to reddening units. I maintain the original angular resolution corresponding to \texttt{NSIDE}=256 and distance binning up to the maximum distance covered (1.95 kpc).

In addition to the E23 3D reddening map, I utilize two external optical catalogs for calibration and independent verification of the final extended map. Specifically, I use a catalog of revised extinctions (visual), stellar radii, and distances for 1.5 million stars. This catalog is based on a combination of APOGEE, GALAH, RAVE, and \textit{Gaia} DR3 data \citep[YK23 further on]{2023ApJS..264...41Y}, and a catalog of classical and type II cepheids identified and characterised in the VISTA Variables in the Vía Láctea (VVV) survey \citep{2019ApJ...883...58D}. This catalog reports extinctions in $K/H$ bands and distance estimate based on variability analysis of VVV data.
Neither of these catalogs contains sufficient number of objects to construct a 3D map of quality comparable to E23 map, and neither was used by E23 to construct their map. However, both include significant number of objects outside of the E23 volume (i.e.$\sim25\%$, see also Fig~\ref{fig:overview}). Note that the extinction measurements presented in YK23 are arguably more reliable than those by \cite{2023MNRAS.524.1855Z} as they are derived based on analysis of high-quality ground-based spectroscopic observations rather than medium-resolution \textit{Gaia BP/RP} spectra. To conclude, these datasets are suitable both for independent verification and calibration both within the original 2\,kpc volume and beyond. I emphasise that these catalogs are not used for construction of the maps but rather only verification and calibration as described in section \ref{sec:calibration}.

\subsection{Dust emission and integrated reddening maps}
\label{sec:input_maps_gnilc}
\begin{figure*}
    \includegraphics[width=0.33\textwidth]{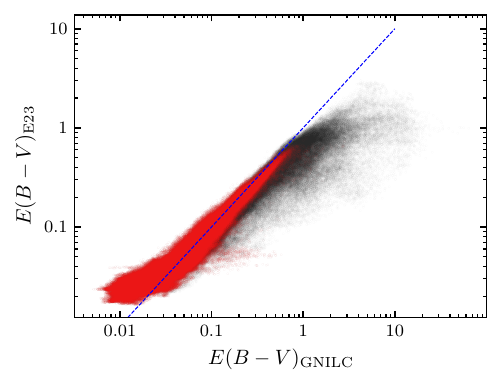}
    \includegraphics[width=0.33\textwidth]{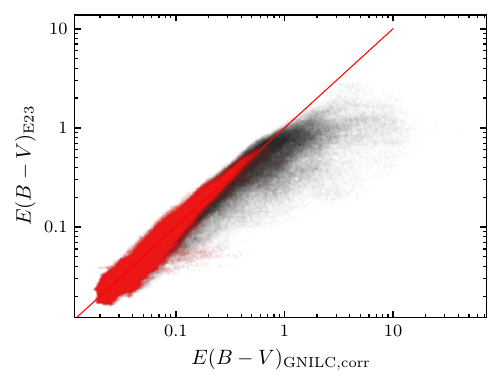}
    \includegraphics[width=0.33\textwidth]{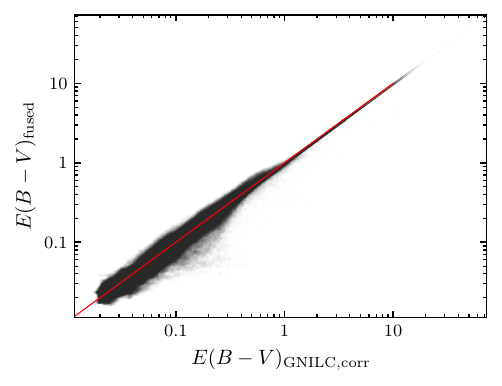}
    \caption{Pixel by pixel comparison of integrated E23 extinction (multiplied by 2.8 to get $A_V$) and \textit{Planck} reddening maps. Red points indicate pixels with $|b|>20^\circ$, i.e. away from the Galactic plane. Left and middle panels shows comparison before and after correction of \textit{GNILC} map respectively. The right panel shows the comparison of the corrected \textit{GNILC} and the final fused maps.}
    \label{fig:E23vsGNILC}
\end{figure*}
Analysis of dust distribution beyond a few kilo-parsecs is not feasible with current optical instrumentation. Therefore, integrated dust emission maps, such as those based on COBE/DIRBE and IRAS/ISSA \citep{1998ApJ...500..525S}, and more recently, \textit{Planck} data \citep{2016A&A...596A.109P}, are commonly used to estimate total absorption in the Galaxy. Unfortunately, constraining the distribution of dust along the line of sight is challenging, so only 2D maps are available. Nevertheless, dust emission maps allow to estimate the fraction of the absorbing material already probed by optical extinction maps, and, what is more important, also how much of it is missing due to limited depth of the optical surveys.

In this work, I use the \textit{Planck/GNILC} reddening map \citep{2016A&A...596A.109P} as input because it appears to give results more consistent with other estimates at least for the extra-Galactic sky \citep{2021MNRAS.503.5351S}. The map is accessible as a reddening $E(B-V)$ map with an angular resolution of $\sim1.7^\prime$ (\texttt{NSIDE}=2048) via the \textit{dustmaps} package \citep{2018JOSS....3..695M}. Considering the lower angular resolution of the E23 map, I start the analysis by down-sampling the \textit{Planck} map to the same resolution as E23 (\texttt{NSIDE=256}). Specifically, I first smooth the original map using a Gaussian kernel with a full width half maximum of $13.74^\prime$, which corresponds to the effective resolution of the E23 map. Then, I use the \texttt{healpy.ud\_grade} function to down-sample the result to \texttt{NSIDE=256}. The smoothing is necessary to avoid artefacts around compact highly absorbed regions close to the borders of larger \texttt{NSIDE=256} pixels and does not affect the results otherwise. Comparison of the up-sampled {NSIDE=256} and original {NSIDE=2048} \textit{Planck} maps allows also to estimate the uncertainty introduced due to the down-sampling procedure. This turns out to be of the order 4\%, although up to factor of two differences are observed for some high-contrast compact regions.

In principle, this is already sufficient to estimate integrated reddening. However, it is important to first verify whether the two maps are consistent and which one produces better results for high Galactic lattitudes. In fact, qualitative comparison of the inner 1.25\,kpc part of E23 with \textit{Planck} and other maps was already discussed by E23, who concluded that there is indeed a very good agreement. E23 also conducted quantitative comparison of their results other 3D maps and extinction values reporter by \cite{2023MNRAS.524.1855Z}. In the later case they found good agreement with differences between 50\,mmag and 4\, mag for inner 1.25\,kpc volume with a dispersion of about 50\,mmag.
Here, I do a similar quantitative comparison of the full E23 (i.e. integrated up to 1.95\,kpc as that is what I intend to use further on), and \textit{GNILC} maps, that is, compare them pixel by pixel. The result is presented in Fig.~\ref{fig:E23vsGNILC}, where E23 reddening map is integrated up to the maximal distance (1.95\,kpc). Although the overall agreement is indeed quite good (at least for intermediate reddenings), several relevant points can be mentioned here:

\begin{figure*}
    \begin{center}
    \includegraphics[width=0.8\textwidth]{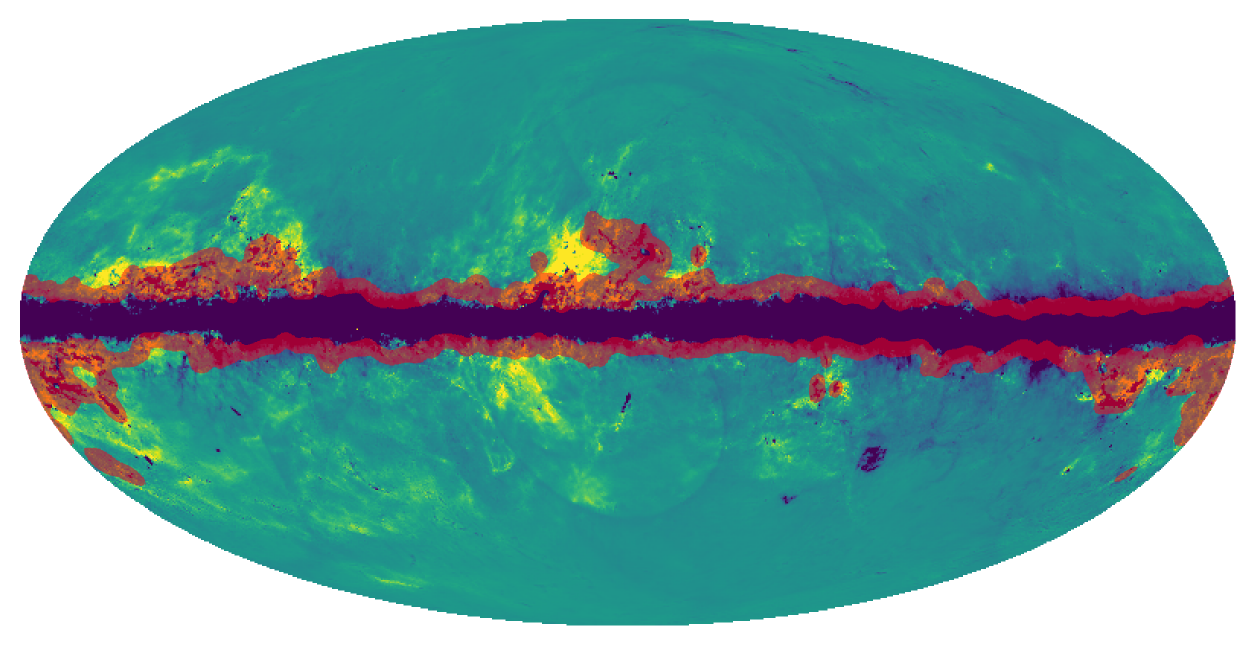}
    \includegraphics[width=0.8\textwidth]{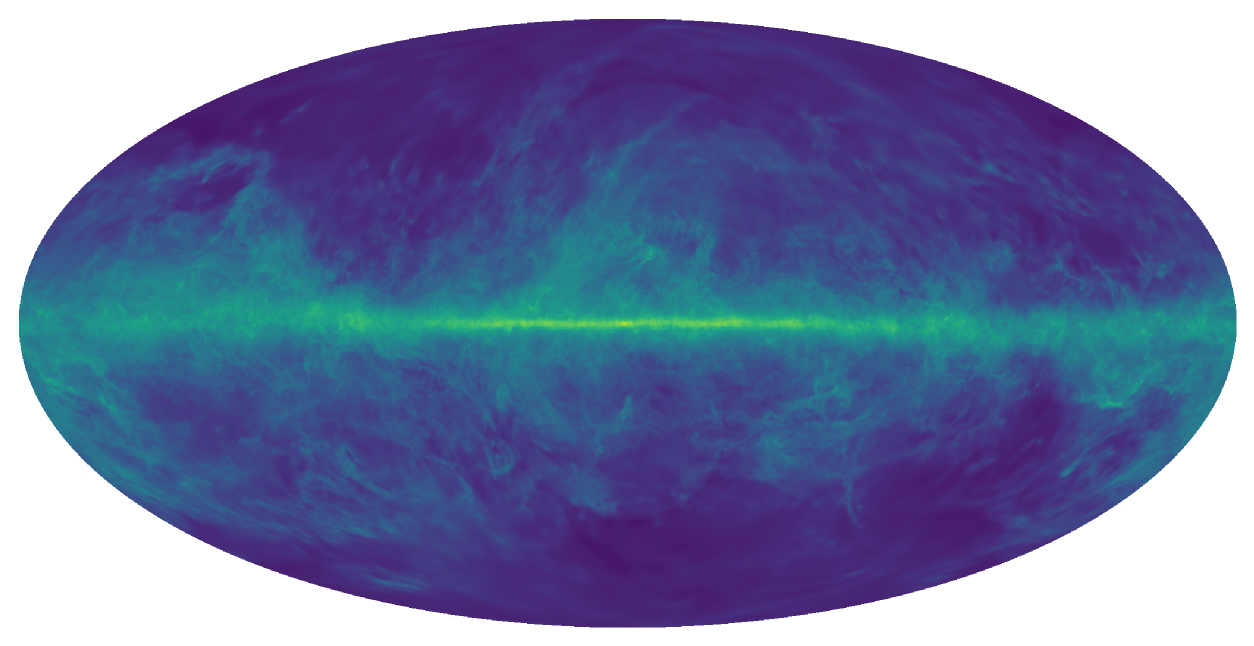}
    \includegraphics[width=0.8\textwidth]{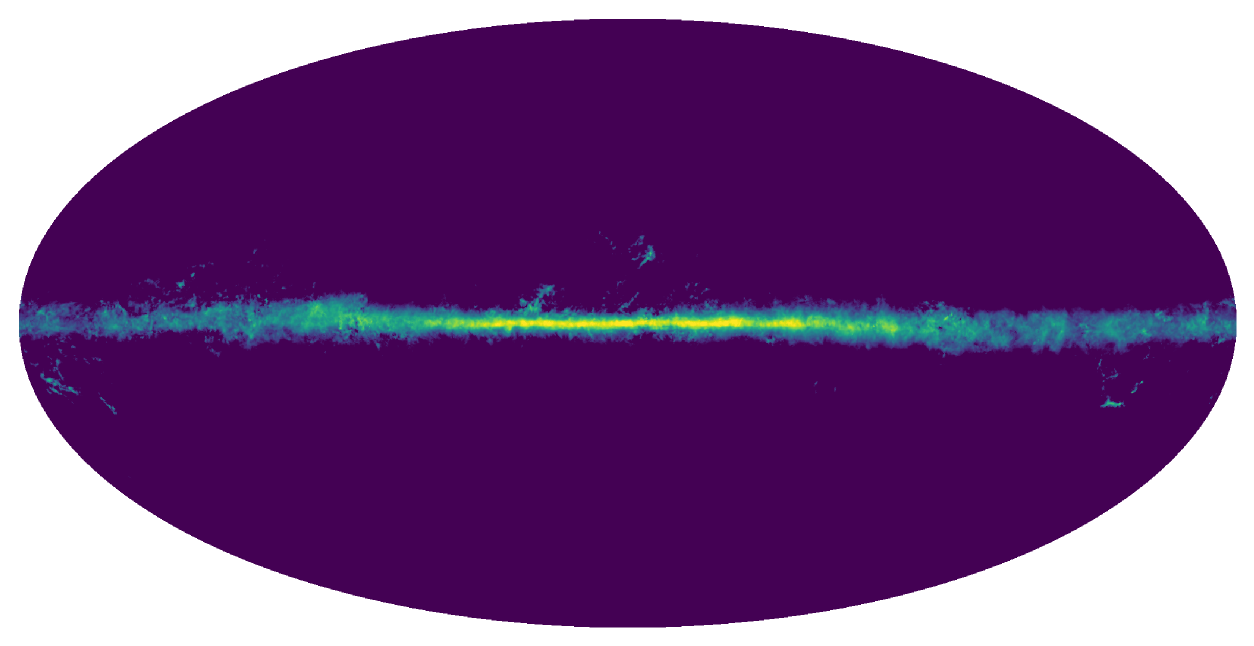}
    \end{center}
    \caption{\textit{Top panel:} difference between integrated E23 and corrected \textit{GNILC} maps (color range is limited to $\pm0.1$ to highlight differences). The red contours show regions where \textit{GNILC} values are used to extend E23 map in the middle panel. 
     \textit{Middle panel:} fused map combining E23 (outside of the contours) and corrected \textit{GNILC} (inside the contours) maps (logarithmic scaling from 0.01 to 100 mag). Note the smooth connection in the transition region.
     \textit{Bottom panel:} difference between final fused map and integrated E23 map (logarithmic scaling from 0.01 to 100 mag, i.e. dark blue coloured sky regions correspond to area where E23 data is used and brighter parts to areas where \textit{GNILC} is used).}
    \label{fig:E23vsGNILC_map}
\end{figure*}

\begin{itemize}
    \item the agreement for intermediate reddenings is generally good, however, E23 map expectedly under predicts
     total reddening in the Galactic plane (by up to 60 magnitudes), i.e. probes only a few percent of total absorbing material. This is unfortunate even if fully expected due to limited depth of the map
    \item \textit{Planck} measures higher reddening also for some denser compact regions also away from the Galactic plane (red outlier points around $E(B-V)_{\rm E23}\sim0.04$. The reason is likely that these features are not fully resolved in E23 due to the limited number of sample stars for these lines of sight
    \item although not obvious from Fig.~\ref{fig:E23vsGNILC}, also for intermediate reddenings the agreement is not perfect, likely due to the uncertainty in the conversion factor from monochromatic extinction to reddening (extinction law $R_V$ and $A_0$ to $A_V$ relation). Therefore, some additional calibration is needed 
    \item enhanced reddening in the direction of Maggellanic clouds in \textit{Planck} maps is obviously associated with dust emission from the clouds themselves, i.e. is not Galactic and does not appear in integrated maps (as it should be)
    \item  on the other hand, \textit{Planck} under predicts reddening for regions with low absorption ($E(B-V)_{\rm E23}\sim0.04$) at high latitudes with the exception of Magellanic clouds. That is, there appears to be an offset between reddening measured in the optical and through dust emission, which also needs to be calibrated. Similar discrepancy was observed by E23 when comparing with \cite{2023MNRAS.524.1855Z} results and its reason is not really clear. In principle, optical extinction maps shall be more sensitive in low-absorption regions where local low density material is much easier detectable than through integrated dust emission. That is, due to the large sheer number of tracer stars, so part of the discrepancy might be attributed to the limited sensitivity of \textit{Planck} maps 
\end{itemize}
The last point is probably the most relevant in context of this work. The fact that E23 predicts for high latitudes higher reddenings than \textit{Planck} implies that despite its comparatively low limiting distance most of the absorbing material within the Galaxy is, in fact, already sampled by E23. This is to be expected as most of the dust is confined to a few hundred parsecs from the Galactic plane, i.e. within volume covered by E23. This fact implies that for high Galactic latitudes E23 already provides full 3D information for dust distribution. However, situation is clearly different for the Galactic plane where the two maps are highly complementary and can be combined.

Before proceeding, however, the maps need to be cross-calibrated. Specifically, I calibrate \textit{Planck} map to match the integrated reddening values predicted by E23 (base map for the analysis) for high Galactic latitudes. To estimate the conversion factor (i.e. correction to standard extinction law), I consider \texttt{HEALPIX} pixels with $|b|>20^\circ$ and $E(B-V)_{\rm GNILC}\le0.3$, i.e. pixels with low to intermediate extinction. These are highlighted with the in Fig.~\ref{fig:E23vsGNILC}. I then fit a model $E(B-V)_{\rm E23} = R_{V,{\rm corr}}\times E(B-V)_{\rm GNILC} + E_0(B-V)$ to these data using the Bayesian inference package \texttt{UltraNest} \citep{2021JOSS....6.3001B} as described in Appendix~\ref{sec:appendix}.
The analysis reveals a constant offset of $\sim0.02$\,mag between the maps, with E23 yielding slightly higher values in low-reddening regions. It also shows that the \textit{GNILC} map needs to be corrected by a factor of $R_{V,{\rm corr}}\sim0.78$ to match E23 reddenings calculated from the raw map as described above. This finding aligns with the results of \cite{2003A&A...408..287D, 2011ApJ...737..103S, 2021MNRAS.503.5351S} where correction factors in range $\sim0.75-0.87$ are reported. The most commonly accepted value appears to be closer to upper end of this range at 0.86 by \cite{2011ApJ...737..103S}, i.e. somewhat higher than found here. I emphasise, however, that exact value of the correction factor is not relevant in context of this work as final calibration will be accomplished using independent absorption tracers as described in section~\ref{sec:calibration} and the goal here is simply to achieve agreement between the values in the two maps. The scatter of the relation is also estimated from the fit at around $\sim13$\% or $\le30$\,mmag within the considered extinction range, which appears to be consistent with estimates by E23.

 The obtained relation was then used to correct the \textit{GNILC} maps to match E23 values. The result of the correction, presented in the right panel of Fig.~\ref{fig:E23vsGNILC}, shows that the agreement is indeed substantially improved. The corrected \textit{GNILC} map can now be used to extend the (integrated) E23 map to the Galactic plane. 

To define what the ``Galactic plane'' actually is, it is useful to inspect the difference between E23 and corrected \textit{GNILC} maps presented in the top panel of Fig.~\ref{fig:E23vsGNILC_map}. As evident from the figure, for most of the sky the agreement is now indeed very good (i.e. the differences are within $\pm0.1$\,mag, which is comparable to estimates of E23 uncertainty itself). There are, however, also some notable differences. First, several apparently local areas (i.e. Taurus, Ophiucus, Orion star forming regions) with intermediate absorption where E23 yields significantly higher reddening values can be identified (perhaps due to limited sensitivity of \textit{Planck} maps). On the other hand, \textit{Planck} also detects emission from Magellanic clouds (i.e. over-predicts reddening there). And of course, there is also the Galactic plane with $|b|\lesssim15^\circ$ where difference reaches tens of magnitudes. When combining the two maps, it is sensible to take advantage of the strengths of each dataset. Specifically, we should retain the higher sensitivity of E23 to local features and its ability to separate local and extra-Galactic features, as well as the sensitivity of \textit{Planck} in highly absorbed areas. 

In practice, this can be achieved by defining a weight mask based on the fit results and difference map presented in Fig.~\ref{fig:E23vsGNILC_map}. 
The mask is defined as follows: E23 values are used for pixels with $E(B-V)_{\rm GNILC, corr}\le0.3$ (approximate break of the correlation in the right panel of Fig.~\ref{fig:corner_e23ebvxgnilcebv}) and pixels where the reddening predicted by E23 is at least 0.05\,mag higher than that predicted by dust maps. This corresponds to a $3\sigma$ deviation from the best-fit relation (shown in Fig.~\ref{fig:corner_e23ebvxgnilcebv}). For the rest of the pixels, i.e. the Galactic plane and some local features \textit{GNILC} values are used. Strict separation would imply, however, that the transition region would not be not fully smooth as there is still some scatter between E23 and \textit{GNILC} values even after correction. 

To address this, I apply a Gaussian kernel with a full width half maximum of $\sim1.8^\circ$ (corresponding to \texttt{NSIDE}=32) to the weight mask in order to smooth the transition, and use smoothed version of the mask to calculate weighted average of E23 and \textit{Planck} maps rather than simply taking pixel values of individual maps. 
The transitional region where the weight deviation from 0 or 1 is larger than 1\% is shown with a red overlay in the top panel of Fig.~\ref{fig:E23vsGNILC_map}. Note that this region is relatively narrow and thus for most of the sky either E23 or \textit{Planck} reddening values are effectively used directly. Finally, I also add a small number ($10^{-5}$ mag, i.e., much lower than the estimated uncertainty of the \textit{GNILC} map) to all pixels in order to ensure that cumulative reddening is always formally increasing when extended 3D map is constructed later on. The result, shown in the middle panel of Fig.~\ref{fig:E23vsGNILC_map}, reveals no obvious artefacts. The transition is smooth, Magellanic clouds are not present while star-forming regions are retained, and the reddening in the Galactic plane is now adequate. I refer to this map as the integrated fused map further on and use it to estimate total cumulative absorption in given direction. Note that meaningful differences compared to E23 map only appear in the Galactic plane, i.e. the integrated fused map is effectively an integrated E23 map extended in the Galactic plane using \textit{Planck} data within the red contours in Fig.~\ref{fig:E23vsGNILC_map}. The differences between them are highlighted in the bottom panel of the same figure and effectively define the areas where the E23 map is not sufficient to probe all of the absorbing material and needs to be extended or corrected.

\begin{figure}[!ht]
    \includegraphics{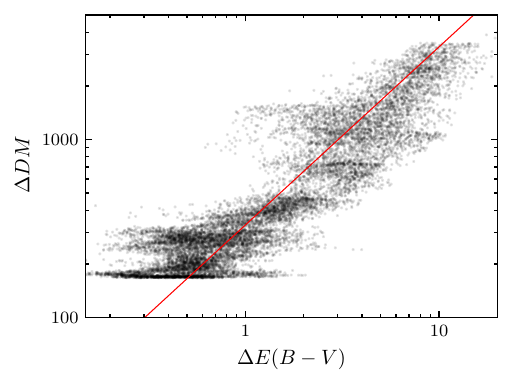}
    \caption{Pixel by pixel comparison of reddening and YMW16 DM model maps integrated between 1.95 and 25\,kpc. The red line shows median ratio between the two quantities.}
    \label{fig:ebvdm}
\end{figure}

\subsection{Dispersion measure models and the third dimension}
\label{sec:input_maps_dm}
The 2D integrated fused map presented above indicates where on the sky the integrated E23 map under predicts reddening and by how much. However, it tells  nothing about the distribution of the missing material along line of sight, which must be estimated by other means. In the simplest case, one could take the difference between the integrated fused map and the E23 map integrated up to 2\,kpc (bottom panel in Fig~\ref{fig:E23vsGNILC_map}), and distribute it uniformly along the line of sight. This approach, in fact, already provides some useful information, but certainly can be improved. For instance, one could use one of the several available 3D HII/CO \citep{2023A&A...671A..54M,2021A&A...655A..64M} maps or dispersion measure (DM) modeling results \citep{2017ApJ...835...29Y}.

Specifically, I opted to use here the YMW16 model of electron density by \cite{2017ApJ...835...29Y}. The motivation for this choice is twofold. First, radio line emission maps necessarily rely on assumptions for the Galactic rotation curve and suffer from serious artefacts in the direction of the Galactic center. In this region, measuring gas velocities and modeling its dynamics are most challenging, and the reddening is highest, so any uncertainties in the modeling translate to large errors in the recovered gas distribution. Second, the \cite{2017ApJ...835...29Y} model includes several large-scale components (Galactic disc, buldge, halo, spiral arms, etc) motivated not only by DM measurements but also by other tracers (stellar clusters, masers, and more) and arguably is easier to interpret when using the final map. For instance, one can associate regions of strong absorption with particular spiral arms, which can be useful for interpretation of the results.

The YMW16 model can be accessed through the \textit{pygedm} package \citep{2021PASA...38...38P}.  Again, before using these results to extend the E23 map it makes sense to compare the two maps and verify that they yield consistent results.  Naturally, the YMW16 lacks all the fine features captured by optical maps. Therefore, it is not sensible to compare it with E23 within the solar vicinity. However, one can compare the cumulative reddening integrated from a larger distance, for instance, from 2\,kpc to infinity. 
I started, therefore, by calculating a set of cumulative \texttt{NSIDE=256} DM maps integrated up to the same set of radial bins as the E23 map (extended to 25\,kpc). The difference between DM integrated up to 25\,kpc and up to 1.95\,kpc can then be compared with the difference between integrated fused map and E23 map.

The result of such a comparison is presented in Fig.~\ref{fig:ebvdm}. Here, $\Delta E(B-V)$ is calculated by subtracting the integrated E23 map from the final integrated fused map and $\Delta DM$ is the difference between YMW16 integrated up to 25\,kpc and up to 1.95\,kpc. Despite the obvious and expected differences (YMW16 contains no small-scale features at these distances), the correlation is still evident (Pearson correlation coefficient $\sim0.86$). This suggests that the YMW16 map does indeed trace also the distribution of colder material responsible for optical extinction over the sky. In the absence of other information, it is thus reasonable to use the YMW16 model to estimate the distribution of material along the line of sight beyond the volume covered by optical maps.  
This can be done by multiplying the difference between integrated fused and E23 maps by normalised cumulative DM from YMW16 model (integrated between 1.95 and 25\,kpc). The result then by definition coincides with the E23 within 1.95\,kpc and with the fused integrated map at infinity while giving some information also for intermediate distances. 
The main question is how accurate this information is, i.e., how well the model actually traces the distribution of the material along the line of sight. This question is discussed in the next section.

\section{Calibration and verification of the extended map}
\label{sec:calibration}
\begin{figure}[!t]
    \centering
    \includegraphics[width=\columnwidth]{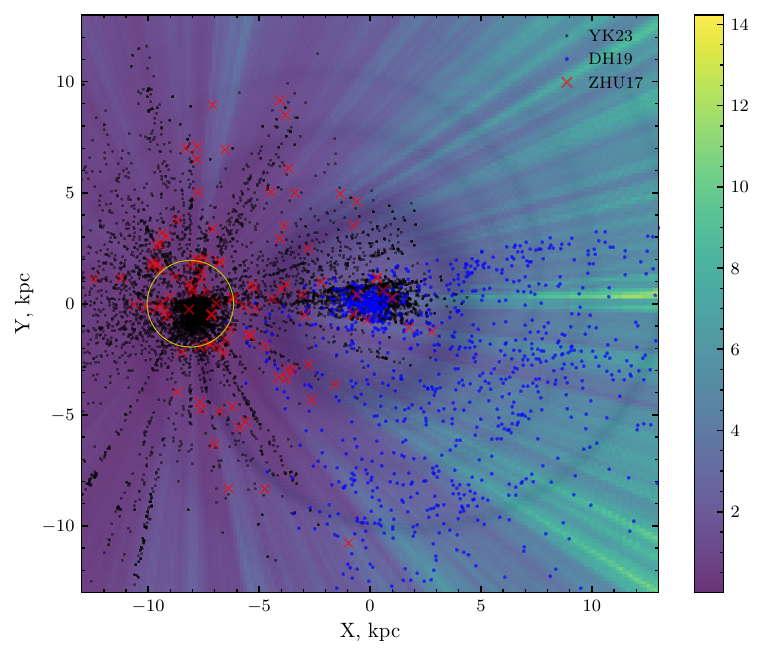}
    \caption{Colour-coded cumulative reddening map in the Galactic plane ($|z|<0.4$\,kpc). Black and blue points indicate location for a subset stars used for verification of the extended map (see main text for detail). The yellow circle indicates volume covered by E23 map. The YMW16 model density used to expand reddening maps is shown to give an idea on location of the spiral arms.}
    \label{fig:overview}
\end{figure}

To evaluate the quality of the extended map and calibrate reddening to standard units (remember the 0.8 factor difference between dust emission and extinction maps discussed in section~\ref{sec:input_maps_gnilc}), I compare the literature optical extinction and X-ray absorption column measurements for a large sample of stars with the values predicted by the extended map. Specifically, for the optical and near infra-red bands I use the catalog of 1.5 million high-quality extinction, stellar radius, and distance measurements by (roughly 25\% of YK23 objects are outside of 2\,kpc E23 volume), and a catalog of Galactic cepheids identified in the VISTA Variables in the Vía Láctea (VVV) survey \citep[see section \ref{sec:input_maps_optical} for more detail]{2019ApJ...883...58D}. The later contains extinction ($K$, and $H$ bands) and distance measurements for around 1000 classical and type II cepheids within VVV footprint, i.e. in the direction of the Galactic center (only one object within 2\,kpc volume).  Finally, I use a compilation of about 100 optical extinction and X-ray absorption column measurements for a sample of supernova-remnants and X-ray binaries compiled by \cite{2017MNRAS.471.3494Z}, referred to as Z17 further on in the text. The locations of these objects within the Galaxy are shown in Fig.~\ref{fig:overview}.

Note that only a fraction of YK23 is plotted to avoid clutter in Solar vicinity. Specifically, I randomly select objects with $|Z|<0.4$\,kpc such that the number of stars roughly constant within logarithmically spaced radial bins. Apparent distribution of black points distribution is thus deliberately biased towards more distant objects to make them more visible and illustrate that there are indeed many objects tracing global stellar density distribution in the Galaxy and located outside of the 2\,kpc volume covered by the original E23 maps. The point is that YK23 is not only useful for re-calibration of E23 maps to standard reddening units, but also for verification of extended map. No additional filtering is applied to the other two smaller catalogs which can be plotted fully.

\begin{figure}[!t]
    \centering
    \includegraphics[width=\columnwidth]{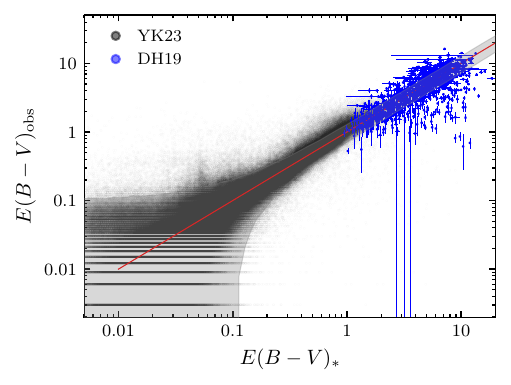}
    \caption{Comparison of reddening values predicted by the extended map obtained here with the values reported by YK23, black points) and \cite{2019ApJ...883...58D} assuming best-fit extinction law for each band (see Table~\ref{table:constants}). The red line shows one to one correspondence and the gray band indicates estimated scatter of the relation.}
    \label{fig:verification_yk23dh19}
\end{figure}

\paragraph{Optical and near-infrared extinction}
\begin{table}[b]
    \caption{Best-fit parameters for linear relation between reddening derived from the final, corrected, 3D map presented here and visual (YK23), $K$-band \citep{2019ApJ...883...58D} extinction values, and X-ray absorption column by Z17. The parameters $R_X$ (conversion factor from $E(B-V)$ to $10^{21}$\,cm$^{-2}$ units), $\sigma_{\rm add}$ (additive error term, either magnitudes or $10^{21}$\,cm$^{-2}$ units) and $\sigma_{\rm rel}$ (multiplicative error term, dimensionless) are found as described in Appendix~\ref{sec:appendix}.}
    \centering
    \label{table:constants}
    \begin{tabular}{lllll}
    \hline
    Band & $R_X$ & $A_K/A_V$ & $\sigma_{\rm add}$ & $\sigma_{\rm rel}$ \\
    \hline
    $V$ & 3.1 & - & 0.1007(4) & 0.277(1) \\
    $K$ & 0.303(8) & 0.098(3) & - & - \\
    X-ray$_{\rm AG89}$  & 7.8(6) & - & 2.4(3) & 0.43(6) \\
    X-ray$_{\rm W00}$  & 11(1) & - & 2.4(6) & 0.4(1) \\ 

    \hline
    \end{tabular}
    
\end{table}
As a first step I calibrate the fused maps obtained as described above to standardised reddening units matching $A_V$ values reported in YK23 for $R_V=3.1$. The motivation is that, as already mentioned, YK23 report actual $A_V$ rather than non-standard monochromatic extinction $A_0$ used by E23, which is not particularly convenient when comparing with the literature. For calibration I use the same Bayesian modeling procedure as for comparison of E23 and \textit{Planck} maps described in Appendix~\ref{sec:appendix}. This allows to estimate both the effective $R_V$ and intrinsic uncertainties for the extended map (again, parametrised as a combination of additive and multiplicative terms). The extended 3D map is then corrected by factor $3.1/R_V$ so that derived reddenings multiplied by 3.1 match the $A_V$ values reported by YK23. Same procedure is effectively applied also to $K$ band. In this case, however, the extinction law is rather uncertain, so $A_V/A_K$ rather than extinction law itself are commonly reported in the literature \citep{2008ApJ...680.1174N}. I thus follow this convention and consider $A_V/A_K$ as a free parameter estimated together $R_V$ and intrinsic scatter of the relation, which is assumed to be the same for both bands. 
The results are presented in Fig.~\ref{fig:verification_yk23dh19}, Fig.~\ref{fig:corner_yk23dh19}, and Table.~\ref{table:constants}. As one can see, the agreement between the reddening values derived from the combined map presented here and the literature is quite good for most objects. Note that some outliers are to be expected as inspection of full resolution \textit{GNILC} map reveals major variations on scales below resolution of E23 map in some areas of the sky. What matters here is, however, average dispersion of the relation, and that is not significantly affected by a handful of outlier points. I use, therefore, estimated conversion factors and uncertainties to convert reddening values from the final 3D reddening map to extinction units for both bands (i.e. $V$ and $K$). The uncertainty for the reddening itself can be then calculated by dividing estimated scatter of $A_V$ by 3.1.
This leads, however, to intrinsic inconsistency with the results of direct comparison of integrated E23 map and \textit{GNILC} maps presented in in section~\ref{sec:input_maps_gnilc} which yields by factor of $\sim3$ lower uncertainties for integrated reddening. To avoid this inconsistency, I assume that 13\% discrepancy between E23 and \textit{Planck} estimated for integrated reddening at high latitudes also applies to the Galactic plane, i.e. use results from Fig.~\ref{fig:corner_e23ebvxgnilcebv} in section \ref{sec:appendix} to estimate the uncertainty of reddening at infinity at any point in the sky. I then calculate the uncertainty for reddening along line of sight as weighted mean of the values derived from extinction calibration and infinity value for given pixel taking fraction of total material on line of sight at given distance as weight. That is, estimated uncertainty for reddening approaches value estimated for infinity for distance objects.
Obviously, a more conservative estimate for uncertainty of the reddening can still be calculated using the values of estimated scatter for the extinction which is also reported.

\paragraph{X-ray absorption column}
In general, the absorption in X-rays is much less explored than in the optical band, making the cross-calibration of the two quantities challenging. Several authors have studied this using independent measurements of both quantities, most commonly in the direction of supernova remnants (SNRs) and X-ray binaries (XRBs), for instance, \cite{1973A&A....26..257R,1975ApJ...198...95G, 1995A&A...293..889P, 2003A&A...408..581V, 2009MNRAS.400.2050G,2011A&A...533A..16W, 2015ApJ...809...66V, 2016ApJ...826...66F,2023A&A...675A.199A,2023A&A...677A.134N}. The range of the conversion factors between $A_V$ and $N_H$ reported in the literature varies from $\sim 1.8-3.3\times10^{21}A_V$ depending on the objects considered, sample size, the methods used to measure $A_V$ and $N_H$, assumed ISM abundances, and so on (Z17). 

In the context of this work, the latter study is the most relevant as Z17 compile a large sample of reasonably reliable (albeit heterogeneous) $N_H$, $A_V$, and crucially, distance estimates for variety of objects. I thus use this database, specifically Table~4 of their paper to calibrate reddening estimates obtained here to X-ray absorption column values reported in the literature. Basically, I follow the same bayesian inference procedure as for other calibrations discussed above substituting the literature $A_V$ estimates used by Z17 with $E(B-V)$ obtained from the 3D maps presented above. 
The aim is again to estimate both the conversion factor between standardised reddening and X-ray absorption column and the intrinsic scatter of the relation. 

Following the approach of Z17, I model separately two subsets of sources where X-ray absorption is reported using two commonly used abundances (i.e., \citealt[or AG89]{1989GeCoA..53..197A}, and \citealt[or W00]{2000ApJ...542..914W}). The absorption column estimates corresponding to W00 case are commonly used in conjunction with more modern \texttt{tbabs} absorption model and thus shall be more reliable, but are only available for a fraction of objects in Z17 sample. The conversion factor from the absorption column to extinction is known to be higher for W00/\texttt{tbabs}, hence the necessity for the separate calibration of the two cases both by Z17 and here.

\begin{figure}
    \includegraphics{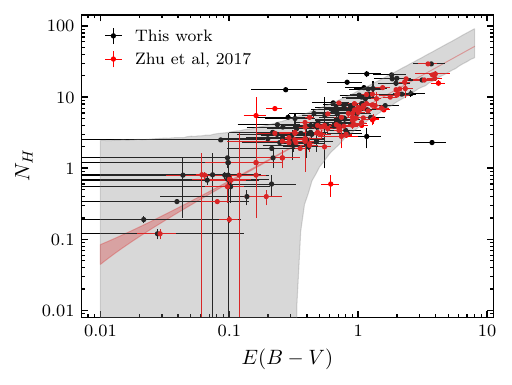}
    \caption{Comparison of reddening derived from the 3D map presented here and X-ray absorption column by Z17 for sample of objects where the later is reported assuming AG89 abundances. The gray band corresponds to $1\sigma$ confidence interval derived for best-fit model. The results by Z17 including model uncertainties (red points and band respectively) are also shown for reference.}
    \label{fig:zhu_vs_ours}
\end{figure}

The results are presented in Figs.~\ref{fig:zhu_vs_ours}, \ref{fig:corner_zhu_ag89}, and Table~\ref{table:constants}. As evident from Fig.~\ref{fig:zhu_vs_ours}, general agreement is very good considering uncertainties of individual points with only several outliers both for our values and original points reported by Z17.
Note, however, that estimated $N_H/A_V$ value I obtain is somewhat higher than those reported in Z17. Same difference, by a factor of $\sim 1/R_{\rm V, corr}\simeq 1.3$, was, in fact, previously observed when discussing the agreement between E23 and \textit{GNILC} values. The likely reason for this discrepancy is that most of the literature visual extinction values are calibrated to \cite{1998ApJ...500..525S} reddenings, which appear to overestimate the true extinction as discussed by \cite{2021MNRAS.503.5351S} and is evident from the comparison of $A_{\rm 0, V}$ values reported by E23,YK23 with \textit{GNILC} maps presented above. It is important to emphasise that the exact relation between the visual extinction and reddening values reported in the literature is beyond the scope of this work. Here, I simply choose YK23 as the baseline for visual extinction units and re-calibrate reddening and X-ray absorption column to match it. This is, however, definitively another factor to keep in mind when comparing results derived from the maps presented here with older literature values (besides the assumed abundances and absorbtion model used).
Finally, I would like to note that the uncertainties for the $N_H/A_V$ factor reported by Z17 are apparently underestimated, i.e. red band in Fig~\ref{fig:zhu_vs_ours} corresponding to $1\sigma$ uncertainty clearly does not cover all the points. I, therefore, stick to more conservative uncertainties derived here when reporting absorption column values in the $N_H$-tool discussed below. Considering the aforementioned advantages of \texttt{tbabs}/W00 abundances, I use the conversion factor derived for this case when reporting X-ray absorption column values in the $N_H$-tool. This has slightly larger estimated uncertainties due to the lower number of objects in calibration sample, i.e. is also a more conservative estimate.

\section{Data access and the 3D-$N_{\rm H}$-tool}
\label{sec:nhtool}
The final calibrated cumulative reddening cube obtained as described above along with the code used to generate and calibrate it are available at 3D-$N_{\rm H}$-tool webpage\footnote{\url{http://astro.uni-tuebingen.de/nh3d}} and \textit{Zenodo}$^{3}$. The cube is stored as a \textit{numpy} archive containing a set of \texttt{HEALPIX} maps in $E(B-V)$ units matching YK23 visual extinction units for $R_V=3.1$ with \textit{NSIDE}=256 for geometrically spaced radial bins taken from E23 map and extended to 25\,kpc (angular resolution of $13.74^\prime$ and radial resolution from 0.2 to 240\,pc). The best-fit values for the conversion factors from $E(B-V)$ to $A_V,A_K$ and X-ray absorption column for AG89 and W00 abundances, as well as their uncertainties are also included in the archive for convenience. The code used to generate the maps and most of the figures in the paper, as well as an example code which can be used to query the maps is also provided. 

In addition, a web-interface has been developed to access the maps, which allows users to query any position on the sky for estimated reddening, extinctions, and X-ray absorption column, including uncertainties. The uncertainties are estimated based on the calibration results of the maps using independent calibration datasets described above. For the absorption column I provide both the estimate obtained through direct conversion of $E(B-V)$ and only taking into the account uncertainties of reddening itself and conversion factor, and separately estimate including estimated scatter of $E(B-V)$ to $N_H$ relation. In addition, I also report for reference the integrated $N_H$ value estimated \textit{HI4PI} maps\footnote{\url{https://heasarc.gsfc.nasa.gov/Images/w3nh/h1_nh_HI4PI.fits}}, i.e. value from HEASARC X-ray absorption column tool. Here the reported value and uncertainty are calculated as mean and standard deviation of the values within the \texttt{HEALPIX} pixel corresponding to the queried position. 

Considering the 3D nature of the maps, uncertainty in distance translates to uncertainty in estimated absorption column. I therefore, make an effort to take this into the account. The distance can thus be specified as part of the query (either a fixed value or range), or automatically looked-up if a unique \textit{Gaia}~DR3 counterpart is identified within 2$^\prime\prime$.
In the later case the geometric priors from \cite{2021AJ....161..147B} are assumed to calculate the absorption column value and uncertainties. Note that identified counterpart is not always reliable, so the user should always check the results and explicitly query \textit{Gaia} DR3 position or specify the distance manually in case of doubt. Users can also access and download cumulative line-of-sight reddening as function of distance using the interactive tool (see i.e. Fig.~\ref{fig:webinterface}). Downloaded curve can then be used to perform more sophisticated analyses without downloading the full cube. The service is accessible at \url{http://astro.uni-tuebingen.de/nh3d}, and some details on implementation as well as a screenshot of the main page are provided in Appendix~\ref{sec:appendix_web}.

\section{Summary and conclusions}
\label{sec:summary}
In this work, I present a new 3D map of Galactic reddening based on the combination of E23 optical extinction maps (which cover high Galactic latitudes within a 2\,kpc local volume), the \emph{Planck/GNILC} \citep{2016A&A...596A.109P} reddening maps (which provide the total absorption column in the Galactic plane), and the YMW16 model \citep{2017ApJ...835...29Y} of electron density (which describes the distribution of material along the line of sight beyond the volume covered by the E23 map). This allows to extend E23 map to distances up to 25\,kpc. The extended map is then calibrated and verified using several independent tracers from the literature, i.e. optical and near-infrared extinction measurements, as well as X-ray absorption column measurements. The main results of this work are as follows:
\begin{itemize}
    \item A pan-Galactic reddening map in $E(B-V)$ units is presented. This map matches the YK23 visual extinction units for $R_V=3.1$. The map is provided as a set of \texttt{HEALPIX} maps with \textit{NSIDE=256} defined for geometrically spaced radial bins taken from the E23 map, extended to 25\,kpc. The angular resolution is thus $13.74^\prime$ radial resolution ranges from 0.2 to 240\,pc. The map is accessible as a raw data cube and via the dedicated 3D-$N_{\rm H}$-tool webpage$^{4}$.
    \item The accuracy of extended map is verified using a sample of approximately $8\times10^{5}$ visual extinction measurements (YK23), around $10^{3}$ $K$-band measurements (\citealt{2019ApJ...883...58D}), and roughly 100 X-ray absorption column measurements (Z17). These measurements include about $4\times10^{5}$ objects outside of the 2\,kpc volume covered by the original E23 cube. 
    \item The extended map appears to agree adequately with the literature, with only a few outliers. More specifically, I estimate the absolute uncertainty to be around 100\,mmag. The relative uncertainty for reddening is estimated to be between 10 and 30\%, depending on the band used for calibration. The estimated conversion factors for visual and $K$-band extinction, X-ray absorption column, and their uncertainties are also provided.
\end{itemize}
The results presented here are primarily intended for practical applications, that is to make the already available 3D absorption maps more accessible and useful for the community. The interactive 3D$N_H$-tool is meant i.e. to serve as a substitute for the HEASARC X-ray absorption column tool$^1$. 

For most of the sky, the results presented here coincide with the input maps (i.e., E23 and \citealt{2016A&A...596A.109P}) scaled by some factor by design. The scaling factor is obtained through independent verification using YK23, \cite{2019ApJ...883...58D}, and Z17 catalogues, i.e. also this case the result presented here have some added value.
The main motivation is, however, to provide at least some estimate of the absorption column for Galactic plane sources. That is particularly important studies of X-ray binaries and similar objects. These are bright enough in the X-ray band to be observed at distances beyond the normal extinction horizon, but are still strongly affected by the absorption which was up to now rather uncertain. The data and code used to obtain the presented results are made publicly available to facilitate further studies and improvements of the map.

\begin{acknowledgements}
    This research has made use of the SIMBAD database VizieR catalogue access tool, operated at CDS, Strasbourg, France. I would like to thank colleagues of the IAAT for years of fruitful discussions and support. In particular Arthur Avakyan for proofreading the manuscript, and Dr. Denis Malyshev help with setting-up of the hosting for the 3D-$N_{\rm H}$-tool and agreeing to keep it running.   
    \end{acknowledgements}

\begin{appendix}
\section{Calibration and verification of the maps}
\label{sec:appendix}
The E23 map, which serves as the basis for this work, is reported in the monochromatic extinction units used by \cite{2023MNRAS.524.1855Z}. For practical use and comparison with other tracers, it is sensible to convert it to $A_V$ visual extinction or standardised $E(B-V)$ reddening units, which match the standard extinction law $R_V=3.1$. This conversion is also necessary if one aims to extend the map using deeper reddening maps that can probe absorption throughout the Galaxy (i.e., the \textit{Planck/GNILC} map used here). From a practical perspective, it makes more sense, however, to first convert the \textit{Planck} map to E23 units, combine them and extend using the YMW16 electron density model, and then calibrate the combined map to match some literature scale for reddening. This is the approach taken here. The steps are thus as follows:
\begin{itemize}
    \item Convert the \textit{Planck/GNILC} map to E23 units and assess the agreement between the two maps.
    \item Combine the two maps to estimate the total Galactic absorption column and extend the E23 map to larger distances using the YMW16 3D electron density model.
    \item Calibrate the resulting pan-Galactic 3D extinction map to match independent literature estimates for visual and near-infrared extinction and X-ray absorption column, and establish scaling factors and uncertainties for the conversion.
\end{itemize}

All the steps above involve comparing two sets of measurements. These could be individual sky pixels when maps are compared, or extinction/reddening measurements for a set of calibration sources. In either case, I fit the corresponding relation with a linear model modified by additional constant in case of \textit{GNILC} to E23 calibration. Estimating uncertainty is more complex. It is clear that besides the statistical uncertainties of the measurements reported in the literature, systematic effects related to the measurements themselves, as well as the limited spatial resolution of the maps, will be present. Therefore, I choose to consider scatter of calibration relations as part of the model and parametrize the remaining systematic uncertainty as a combination of additive and multiplicative terms. These terms are considered as independent parameters. I then conduct maximum likelihood fitting using the Bayesian inference package \texttt{UltraNest} \citep{2021JOSS....6.3001B}. Specifically, I use a likelihood function defined as:
$$
\mathcal{L} = \prod_{i=1}^{N} \frac{1}{\sqrt{2\pi}\sigma_i} \exp\left(-\frac{1}{2}\left(\frac{y_i - (a x_i + b)}{\sigma_i}\right)^2\right)
$$
where $\sigma_i=\sqrt{\sigma_{\rm add}^2 + (\sigma_{\rm rel}y_i)^2}$. Here, $y_i$ and $x_i$ are the measurements and predictions, respectively, and $a$ and $b$ are the parameters of the linear model. The offset $b$ is only used for the cross-calibration of \textit{Planck} and E23 maps. 
The posterior probability distributions and the Bayesian
evidence are then derived with the nested sampling Monte Carlo algorithm
MLFriends \citep{2016S&C....26..383B,2019PASP..131j8005B} using the
\texttt{UltraNest}\footnote{\url{https://johannesbuchner.github.io/UltraNest/}} package \citep{2021JOSS....6.3001B}. Broad flat uniform priors are used for all parameters. The posterior distribution is derived using a nested sampling algorithm with 2000 live points. 

\begin{figure*}
    \centering
    \includegraphics[width=\textwidth]{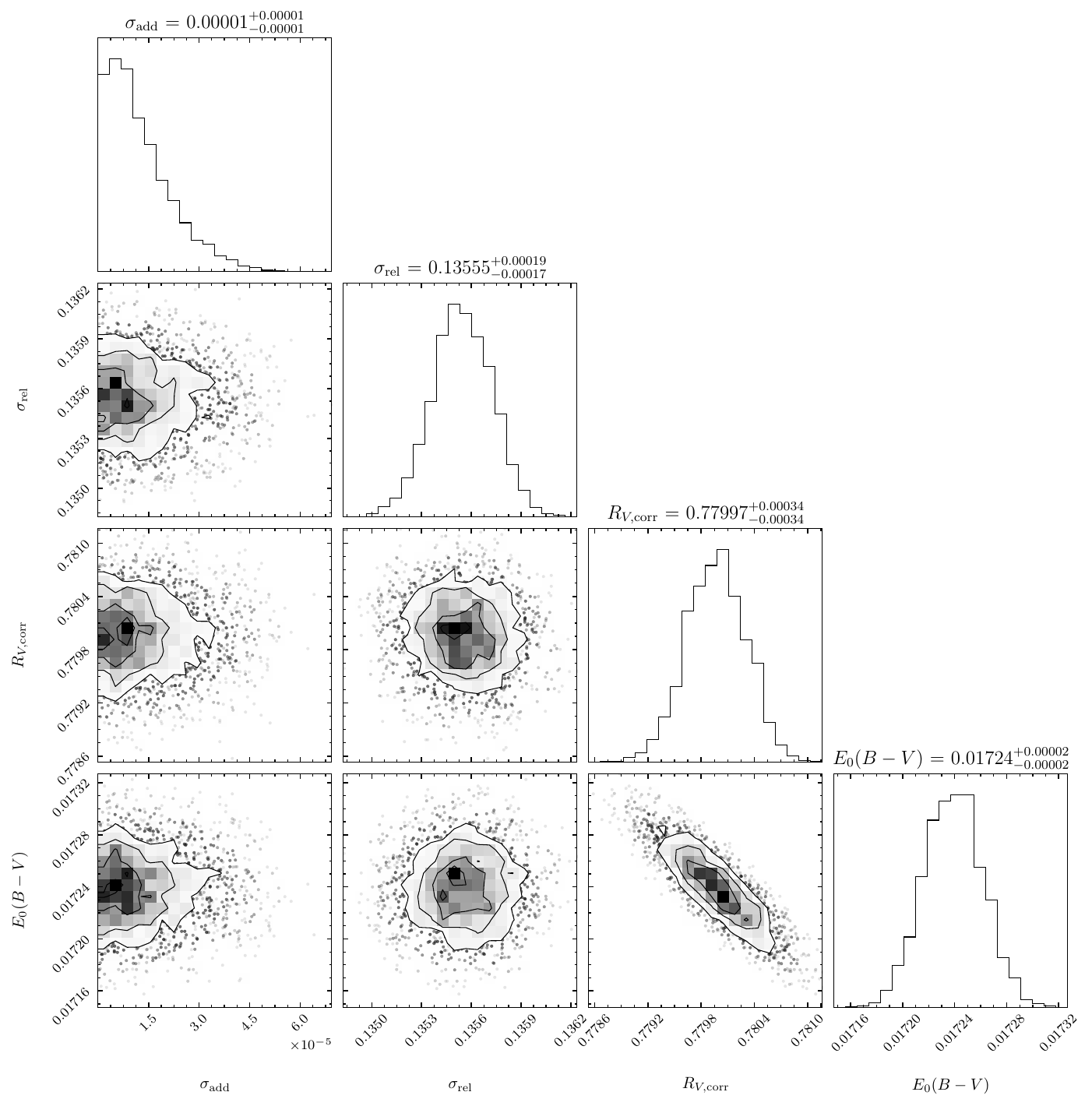}
    \caption{Posterior distribution for parameters of linear relation between E23 and \textit{Planck} reddening.}
    \label{fig:corner_e23ebvxgnilcebv}
\end{figure*}

For the cross-calibration of E23 and \textit{Planck} maps, I disregard the uncertainties of individual pixels because the goal is to estimate the overall agreement between the two maps. Therefore, I simply use likelihood definition above summing over the sky pixels. The posterior distribution of the model parameters is presented in Fig~\ref{fig:corner_e23ebvxgnilcebv} and discussed in the main text.

\begin{figure*}
    \centering
    \includegraphics[width=\textwidth]{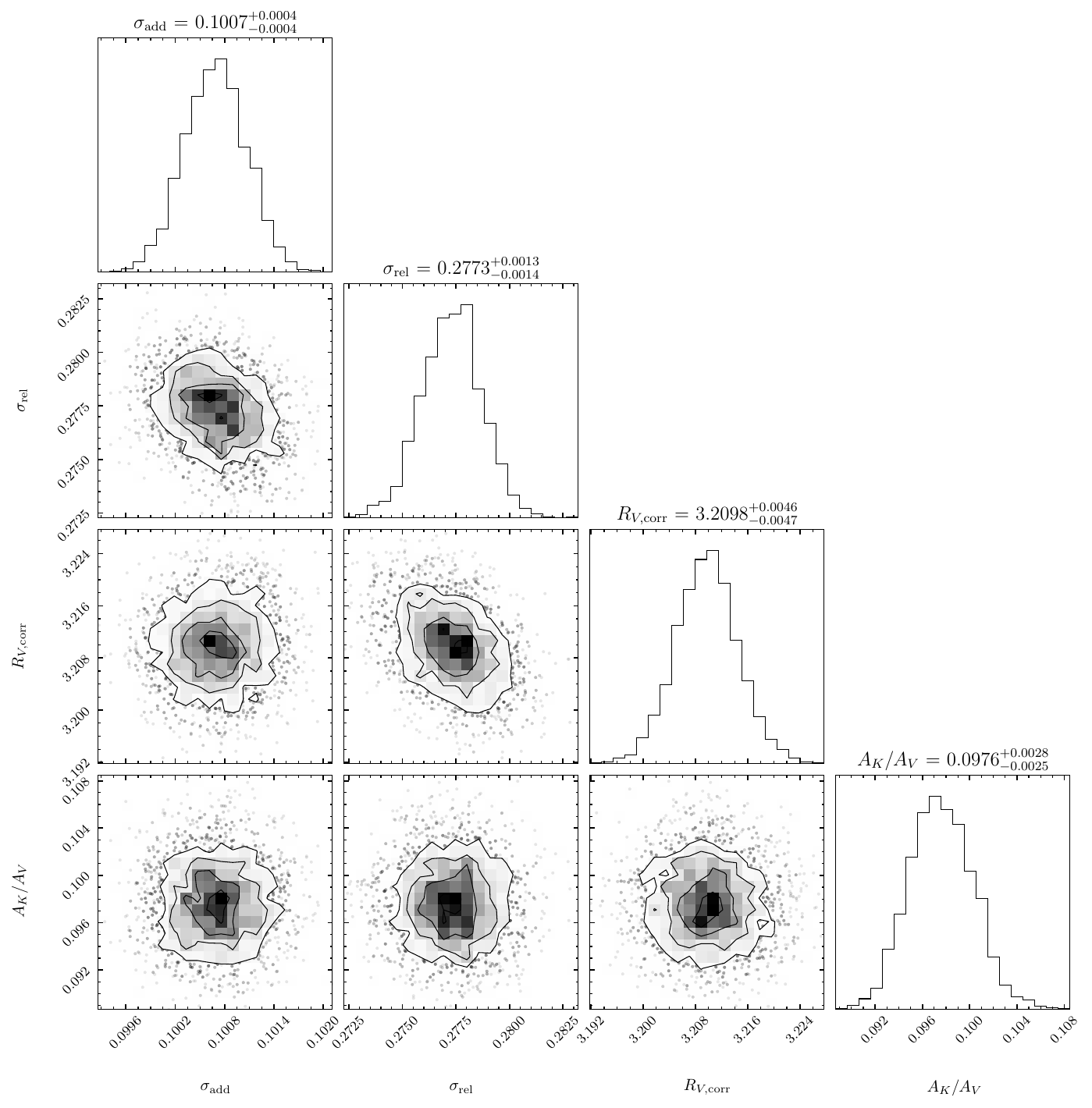}
    \caption{Posterior distribution for parameters of linear relation between reddening derived from the 3D map presented here and visual (YK23) or $K$-band \citep{2019ApJ...883...58D} extinction values reported in the literature.}
    \label{fig:corner_yk23dh19}
\end{figure*}

To calibrate the combined map to match the standardised reddening scaled to visual extinction and the standard extinction law $R_V=3.1$, I adopt the same approach comparing the reddening values predicted by the extended map with the values reported by YK23 and \cite{2019ApJ...883...58D}. That is, the reddening from the map is considered as $x_i$ and the extinction measurements of individual objects as $y_i$.  Since both variables are subject to uncertainties, I do not use the reported extinction values directly. Instead, I randomly sample several realisations (10 samples for YK23 $A_V$ measurements and 100 samples for \cite{2019ApJ...883...58D} $A_K$ measurements) based on the reported mean extinction and its uncertainty for each object. The uncertainty for the reddening in our map is calculated using the results of the comparison of E23 and \textit{Planck}, as discussed above. 
The likelihood is then calculated by summing the mean likelihood values across simulation realisations for individual objects when carrying out the inference. The modeling is done jointly for $V$ and $K$ bands, and the $A_V/A_K$ ratio is considered as an additional free parameter. The intrinsic scatter of the relation is thus assumed to be the same for both bands which is reasonable as uncertainties of individual measurements should be already taking into the account when calculating likelihood function. I opted for this approach, rather than an independent estimate of $R_K$, to facilitate comparison of $A_V/A_K$ with the literature values \citep{2008ApJ...680.1174N}. The posterior distribution of model parameters is presented in Fig~\ref{fig:corner_yk23dh19}. As evident from the figure and discussed in the main text, the derived scaling factors and uncertainties are indeed in agreement with the literature values.

\begin{figure*}
    \centering
    \includegraphics[width=0.49\textwidth]{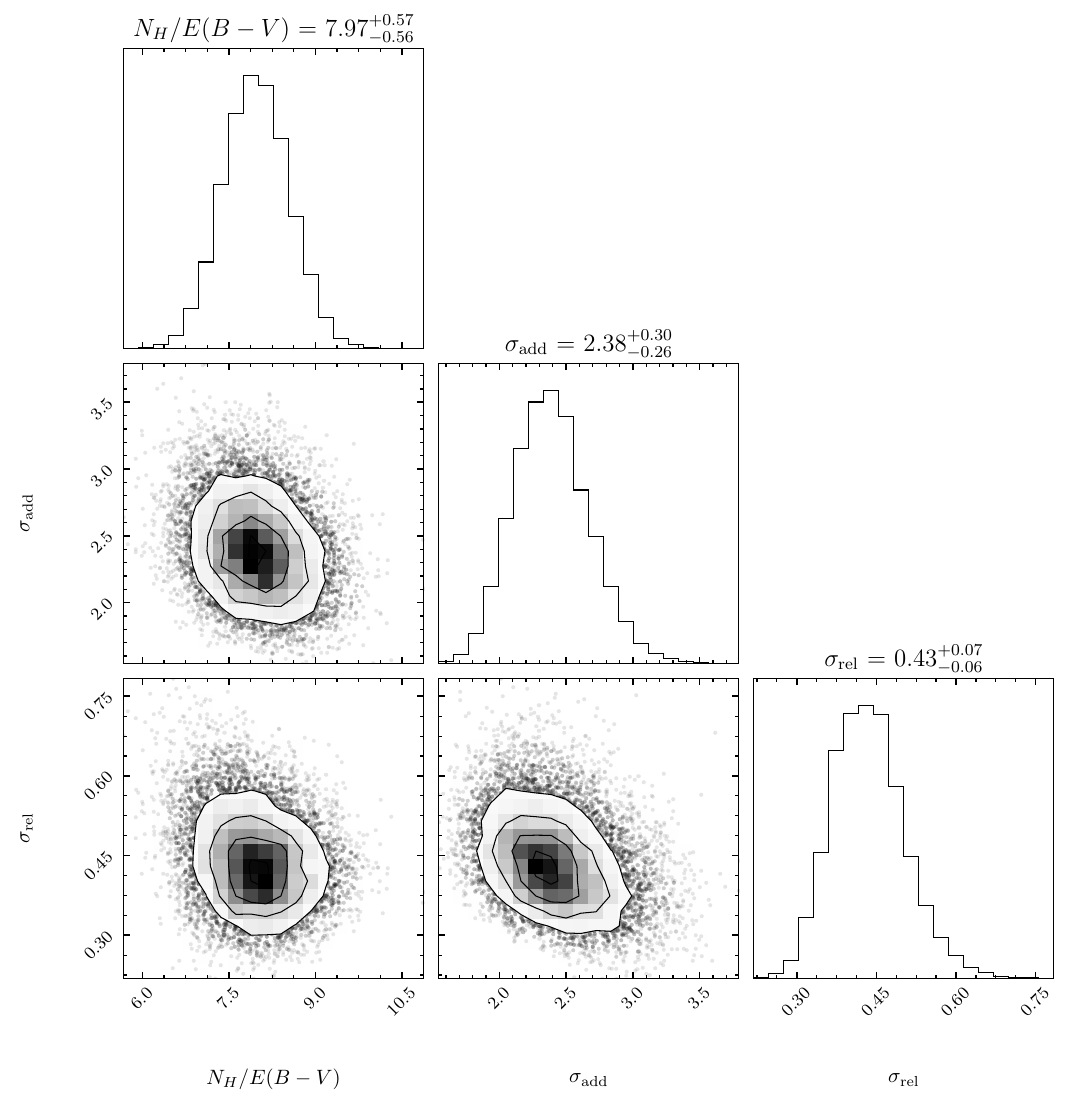}
    \includegraphics[width=0.49\textwidth]{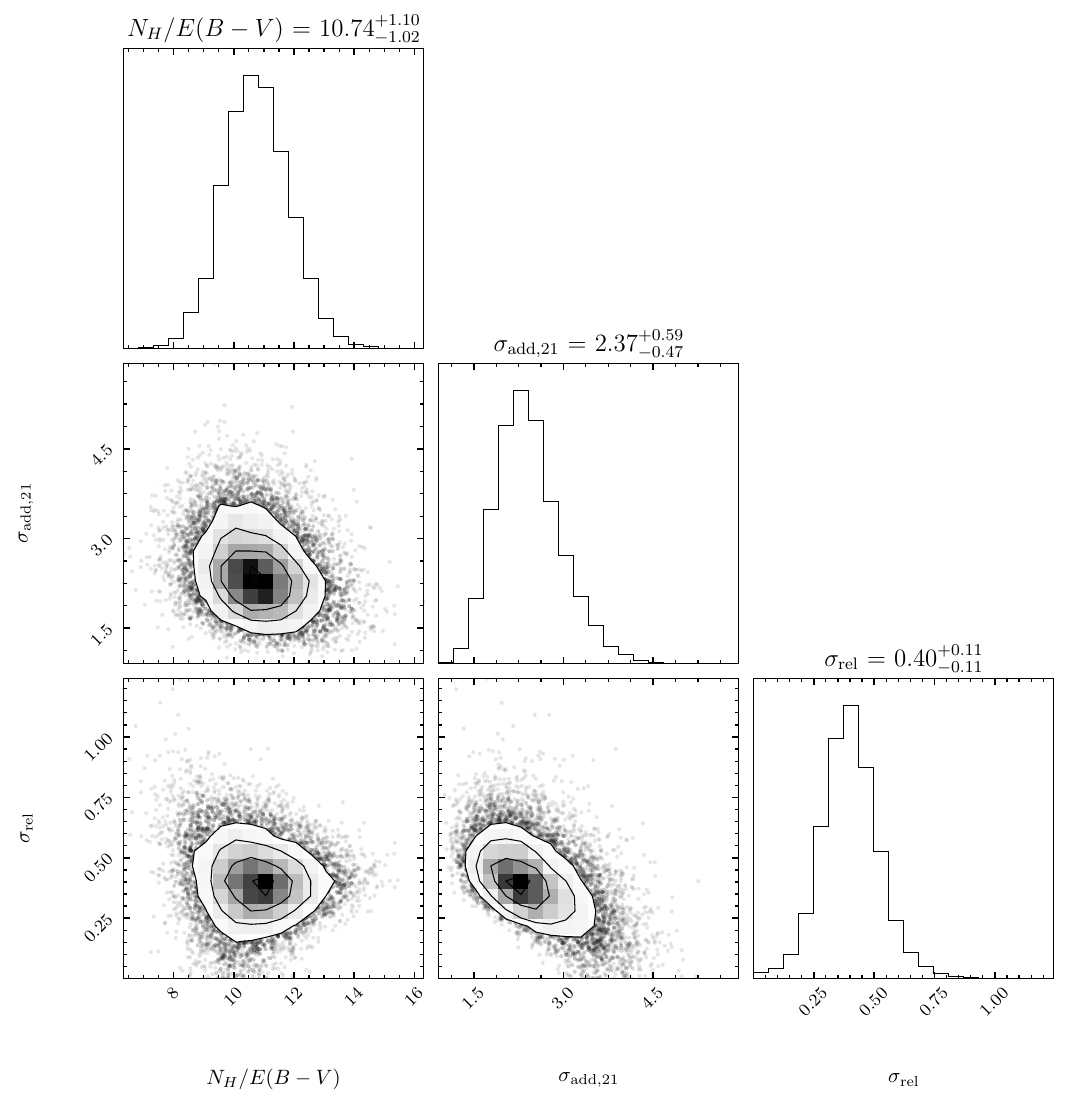}
    \caption{Posterior distribution for parameters of linear relation between reddening derived from the 3D map presented here and X-ray absorption column by Z17 for AG89 (left) and W00 (right) abundances.}
    \label{fig:corner_zhu_ag89}
\end{figure*}

The same procedure is adopted to calibrate the combined map to match the X-ray absorption column measurements by Z17 for AG89 and W00 abundances (independently). The posterior distribution of model parameters is presented in Fig~\ref{fig:corner_zhu_ag89}. Given that column depth values reported for W00 abundances are derived using the \texttt{tbabs} model \citep{2000ApJ...542..914W}, which is considered more reliable, I use the scaling obtained for the W00 sub-sample in Z17 as the default scaling for the X-ray absorption column. The final values for all conversion factors are presented in Table~\ref{table:constants} and are used to calculate relevant quantities in the 3D-$N_{\rm H}$-tool.

\section{3D-$N_{\rm H}$-tool implementation}
\label{sec:appendix_web}
To make the results more accessible to a general user, I have also implemented a web interface, 3D-$N_{\rm H}$-tool, to access the maps. The interface is implemented using python framework \textit{Bokeh} framework\footnote{\url{https://bokeh.org}} and directly interfaces the \textit{numpy} archive containing the maps. The interface allows users to query any position on the sky for estimated reddening, extinctions in $A_{V}$ and $A_K$ bands, and X-ray absorption column (W00 abundances), including their uncertainties estimated as described above. 

\begin{figure*}[!t]
    \centering
    \includegraphics[width=\textwidth]{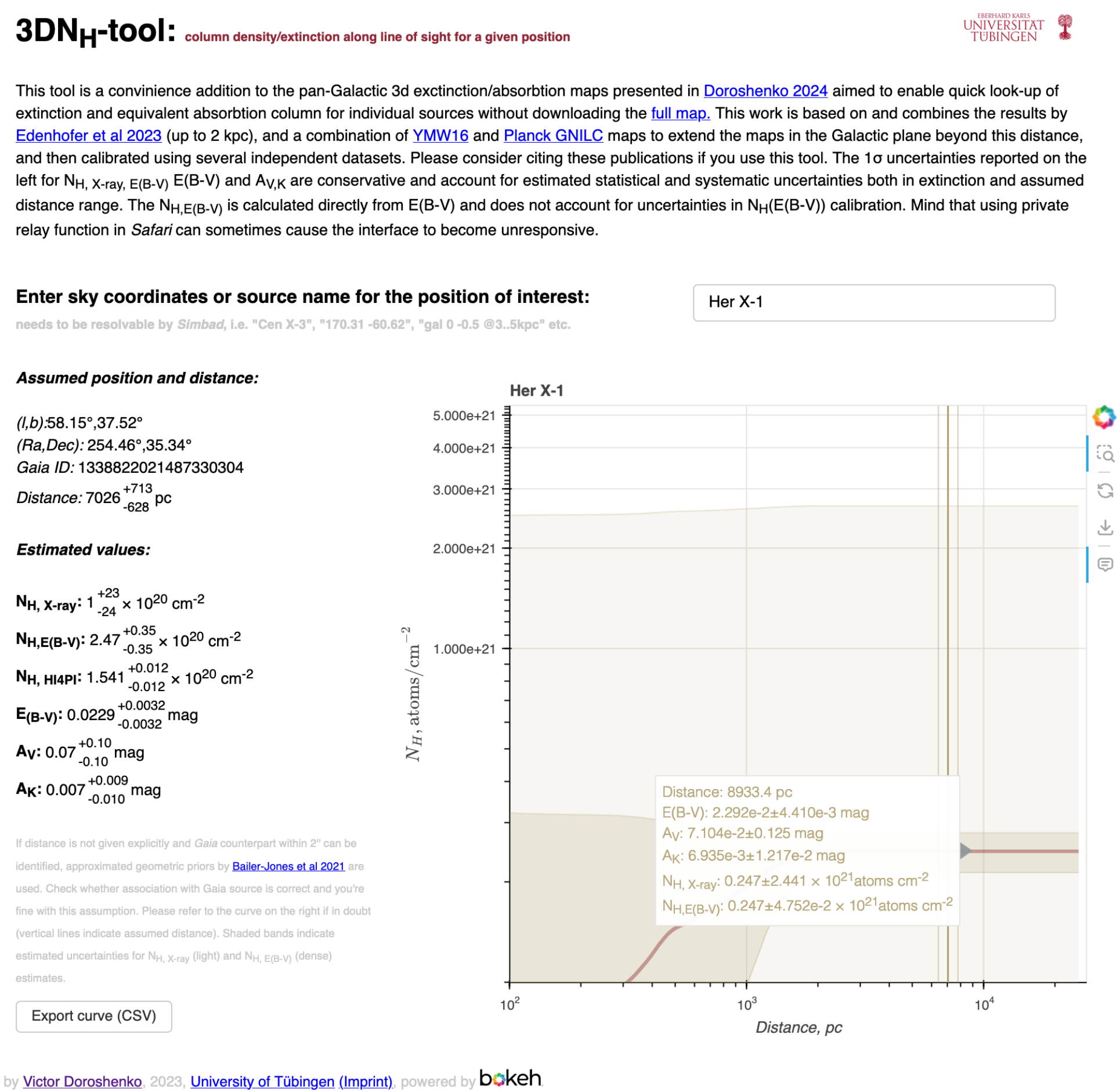}
    \caption{Screenshot of the main page of the 3D-$N_{\rm H}$-tool web-interface to access the 3D reddening map presented here. The left panel shows identified \textit{Gaia} counterpart used to estimate distance to the queried object, and estimated X-ray absorption, reddening and $A_V$ and $A_K$ extinction. For X-ray absorption column two estimates are provided: the more conservative one is based on calibration of the map using independent measurements of X-ray absorption column, while the other one is just reddening scaled by constant factor. The right panel shows line-of-sight r X-ray absorption column (both conservative and scaled reddening estimates) as function of distance. Relevant values for given distance are displayed in the tooltip in the interactive plot on the right. The data can be downloaded in tabular form using the button in the bottom left corner.}
    \label{fig:webinterface}
\end{figure*}

The position can be specified as Simbad-resolvable name or coordinates in flexible format, including distance or distance range (just append something like \texttt{100pc} or \texttt{@1..2kpc} to the query to consider distance of 100\,pc or range 1-2\,kpc). For objects with a unique \textit{Gaia}~DR3 counterpart, an attempt will be made to identify that and use geometric priors from \cite{2021AJ....161..147B} for distance. If distance is not specified and no \textit{Gaia}~DR3 counterpart is found, the integrated map will be queried (i.e. 25\,kpc assumed for distance). In all three cases the values and uncertainties displayed in the left panel of the interface are calculated taking into the account specified distance or distance range (i.e. either constant value, uniform priors within specified range or \cite{2021AJ....161..147B} geometric priors).  Users can also explore line-of-sight dependence of the parameters interactively, save current view of interactive plot, and download displayed dataset including line-of-sight reddening, extinctions and X-ray absorption column values (and uncertainties) in tabular form if more sophisticated analyses if needed. The service is accessible at \url{http://astro.uni-tuebingen.de/nh3d}. A screenshot of the main page is presented in Fig.~\ref{fig:webinterface}.

\end{appendix}

\bibliographystyle{aa}
\bibliography{biblio}

\begin{thebibliography}{55}
\expandafter\ifx\csname natexlab\endcsname\relax\def\natexlab#1{#1}\fi

\bibitem[{{Anders} \& {Grevesse}(1989)}]{1989GeCoA..53..197A}
{Anders}, E. \& {Grevesse}, N. 1989, \gca, 53, 197

\bibitem[{{Anders} {et~al.}(2019){Anders}, {Khalatyan}, {Chiappini}, {Queiroz}, {Santiago}, {Jordi}, {Girardi}, {Brown}, {Matijevi{\v{c}}}, {Monari}, \& et~al.}]{2019A&A...628A..94A}
{Anders}, F., {Khalatyan}, A., {Chiappini}, C., {et~al.} 2019, \aap, 628, A94

\bibitem[{{Avakyan} {et~al.}(2023){Avakyan}, {Neumann}, {Zainab}, {Doroshenko}, {Wilms}, \& {Santangelo}}]{2023A&A...675A.199A}
{Avakyan}, A., {Neumann}, M., {Zainab}, A., {et~al.} 2023, \aap, 675, A199

\bibitem[{{Bailer-Jones} {et~al.}(2021){Bailer-Jones}, {Rybizki}, {Fouesneau}, {Demleitner}, \& {Andrae}}]{2021AJ....161..147B}
{Bailer-Jones}, C.~A.~L., {Rybizki}, J., {Fouesneau}, M., {Demleitner}, M., \& {Andrae}, R. 2021, \aj, 161, 147

\bibitem[{{Buchner}(2016)}]{2016S&C....26..383B}
{Buchner}, J. 2016, Statistics and Computing, 26, 383

\bibitem[{{Buchner}(2019)}]{2019PASP..131j8005B}
{Buchner}, J. 2019, \pasp, 131, 108005

\bibitem[{{Buchner}(2021)}]{2021JOSS....6.3001B}
{Buchner}, J. 2021, The Journal of Open Source Software, 6, 3001

\bibitem[{{Chiang}(2023)}]{2023ApJ...958..118C}
{Chiang}, Y.-K. 2023, \apj, 958, 118

\bibitem[{{Cordes} \& {Lazio}(2002)}]{2002astro.ph..7156C}
{Cordes}, J.~M. \& {Lazio}, T.~J.~W. 2002, arXiv e-prints, astro

\bibitem[{{Dame} {et~al.}(2001){Dame}, {Hartmann}, \& {Thaddeus}}]{2001ApJ...547..792D}
{Dame}, T.~M., {Hartmann}, D., \& {Thaddeus}, P. 2001, \apj, 547, 792

\bibitem[{{D{\'e}k{\'a}ny} {et~al.}(2019){D{\'e}k{\'a}ny}, {Hajdu}, {Grebel}, \& {Catelan}}]{2019ApJ...883...58D}
{D{\'e}k{\'a}ny}, I., {Hajdu}, G., {Grebel}, E.~K., \& {Catelan}, M. 2019, \apj, 883, 58

\bibitem[{{Dickey} \& {Lockman}(1990)}]{1990ARA&A..28..215D}
{Dickey}, J.~M. \& {Lockman}, F.~J. 1990, \araa, 28, 215

\bibitem[{{Drimmel} {et~al.}(2003){Drimmel}, {Cabrera-Lavers}, \& {L{\'o}pez-Corredoira}}]{2003A&A...409..205D}
{Drimmel}, R., {Cabrera-Lavers}, A., \& {L{\'o}pez-Corredoira}, M. 2003, \aap, 409, 205

\bibitem[{{Dutra} {et~al.}(2003){Dutra}, {Ahumada}, {Clari{\'a}}, {Bica}, \& {Barbuy}}]{2003A&A...408..287D}
{Dutra}, C.~M., {Ahumada}, A.~V., {Clari{\'a}}, J.~J., {Bica}, E., \& {Barbuy}, B. 2003, \aap, 408, 287

\bibitem[{{Edenhofer} {et~al.}(2023){Edenhofer}, {Zucker}, {Frank}, {Saydjari}, {Speagle}, {Finkbeiner}, \& {En{\ss}lin}}]{2023arXiv230801295E}
{Edenhofer}, G., {Zucker}, C., {Frank}, P., {et~al.} 2023, arXiv e-prints, arXiv:2308.01295

\bibitem[{{Foight} {et~al.}(2016){Foight}, {G{\"u}ver}, {{\"O}zel}, \& {Slane}}]{2016ApJ...826...66F}
{Foight}, D.~R., {G{\"u}ver}, T., {{\"O}zel}, F., \& {Slane}, P.~O. 2016, \apj, 826, 66

\bibitem[{{Gaia Collaboration} {et~al.}(2021){Gaia Collaboration}, {Brown}, {Vallenari}, {Prusti}, {de Bruijne}, {Babusiaux}, {Biermann}, {Creevey}, {Evans}, {Eyer}, \& et~al.}]{2021A&A...649A...1G}
{Gaia Collaboration}, {Brown}, A.~G.~A., {Vallenari}, A., {et~al.} 2021, \aap, 649, A1

\bibitem[{{Gorenstein}(1975)}]{1975ApJ...198...95G}
{Gorenstein}, P. 1975, \apj, 198, 95

\bibitem[{{G{\'o}rski} {et~al.}(2005){G{\'o}rski}, {Hivon}, {Banday}, {Wandelt}, {Hansen}, {Reinecke}, \& {Bartelmann}}]{2005ApJ...622..759G}
{G{\'o}rski}, K.~M., {Hivon}, E., {Banday}, A.~J., {et~al.} 2005, \apj, 622, 759

\bibitem[{{Green}(2018)}]{2018JOSS....3..695M}
{Green}, G. 2018, The Journal of Open Source Software, 3, 695

\bibitem[{{Green} {et~al.}(2019){Green}, {Schlafly}, {Zucker}, {Speagle}, \& {Finkbeiner}}]{2019ApJ...887...93G}
{Green}, G.~M., {Schlafly}, E., {Zucker}, C., {Speagle}, J.~S., \& {Finkbeiner}, D. 2019, \apj, 887, 93

\bibitem[{{G{\"u}ver} \& {{\"O}zel}(2009)}]{2009MNRAS.400.2050G}
{G{\"u}ver}, T. \& {{\"O}zel}, F. 2009, \mnras, 400, 2050

\bibitem[{{He} {et~al.}(2013){He}, {Ng}, \& {Kaspi}}]{2013ApJ...768...64H}
{He}, C., {Ng}, C.~Y., \& {Kaspi}, V.~M. 2013, \apj, 768, 64

\bibitem[{{HI4PI Collaboration} {et~al.}(2016){HI4PI Collaboration}, {Ben Bekhti}, {Fl{\"o}er}, {Keller}, {Kerp}, {Lenz}, {Winkel}, {Bailin}, {Calabretta}, {Dedes}, \& et~al.}]{2016A&A...594A.116H}
{HI4PI Collaboration}, {Ben Bekhti}, N., {Fl{\"o}er}, L., {et~al.} 2016, \aap, 594, A116

\bibitem[{{Manchester} {et~al.}(2005){Manchester}, {Hobbs}, {Teoh}, \& {Hobbs}}]{2005AJ....129.1993M}
{Manchester}, R.~N., {Hobbs}, G.~B., {Teoh}, A., \& {Hobbs}, M. 2005, \aj, 129, 1993

\bibitem[{{Marshall} {et~al.}(2006){Marshall}, {Robin}, {Reyl{\'e}}, {Schultheis}, \& {Picaud}}]{2006A&A...453..635M}
{Marshall}, D.~J., {Robin}, A.~C., {Reyl{\'e}}, C., {Schultheis}, M., \& {Picaud}, S. 2006, \aap, 453, 635

\bibitem[{{Mertsch} \& {Phan}(2023)}]{2023A&A...671A..54M}
{Mertsch}, P. \& {Phan}, V.~H.~M. 2023, \aap, 671, A54

\bibitem[{{Mertsch} \& {Vittino}(2021)}]{2021A&A...655A..64M}
{Mertsch}, P. \& {Vittino}, A. 2021, \aap, 655, A64

\bibitem[{{Minniti}(2023)}]{2023arXiv230612894M}
{Minniti}, D. 2023, arXiv e-prints, arXiv:2306.12894

\bibitem[{{Neckel} \& {Klare}(1980)}]{1980A&AS...42..251N}
{Neckel}, T. \& {Klare}, G. 1980, \aaps, 42, 251

\bibitem[{{Neumann} {et~al.}(2023){Neumann}, {Avakyan}, {Doroshenko}, \& {Santangelo}}]{2023A&A...677A.134N}
{Neumann}, M., {Avakyan}, A., {Doroshenko}, V., \& {Santangelo}, A. 2023, \aap, 677, A134

\bibitem[{{Nishiyama} {et~al.}(2008){Nishiyama}, {Nagata}, {Tamura}, {Kandori}, {Hatano}, {Sato}, \& {Sugitani}}]{2008ApJ...680.1174N}
{Nishiyama}, S., {Nagata}, T., {Tamura}, M., {et~al.} 2008, \apj, 680, 1174

\bibitem[{{Planck Collaboration} {et~al.}(2016){Planck Collaboration}, {Aghanim}, {Ashdown}, {Aumont}, {Baccigalupi}, {Ballardini}, {Banday}, {Barreiro}, {Bartolo}, {Basak}, \& et~al.}]{2016A&A...596A.109P}
{Planck Collaboration}, {Aghanim}, N., {Ashdown}, M., {et~al.} 2016, \aap, 596, A109

\bibitem[{{Pohl} {et~al.}(2008){Pohl}, {Englmaier}, \& {Bissantz}}]{2008ApJ...677..283P}
{Pohl}, M., {Englmaier}, P., \& {Bissantz}, N. 2008, \apj, 677, 283

\bibitem[{{Predehl} \& {Schmitt}(1995)}]{1995A&A...293..889P}
{Predehl}, P. \& {Schmitt}, J.~H.~M.~M. 1995, \aap, 293, 889

\bibitem[{{Price} {et~al.}(2021){Price}, {Flynn}, \& {Deller}}]{2021PASA...38...38P}
{Price}, D.~C., {Flynn}, C., \& {Deller}, A. 2021, \pasa, 38, e038

\bibitem[{{Queiroz} {et~al.}(2018){Queiroz}, {Anders}, {Santiago}, {Chiappini}, {Steinmetz}, {Dal Ponte}, {Stassun}, {da Costa}, {Maia}, {Crestani}, \& et~al.}]{2018MNRAS.476.2556Q}
{Queiroz}, A.~B.~A., {Anders}, F., {Santiago}, B.~X., {et~al.} 2018, \mnras, 476, 2556

\bibitem[{{Reina} \& {Tarenghi}(1973)}]{1973A&A....26..257R}
{Reina}, C. \& {Tarenghi}, M. 1973, \aap, 26, 257

\bibitem[{{Sale} {et~al.}(2014){Sale}, {Drew}, {Barentsen}, {Farnhill}, {Raddi}, {Barlow}, {Eisl{\"o}ffel}, {Vink}, {Rodr{\'\i}guez-Gil}, \& {Wright}}]{2014MNRAS.443.2907S}
{Sale}, S.~E., {Drew}, J.~E., {Barentsen}, G., {et~al.} 2014, \mnras, 443, 2907

\bibitem[{{Schlafly} \& {Finkbeiner}(2011)}]{2011ApJ...737..103S}
{Schlafly}, E.~F. \& {Finkbeiner}, D.~P. 2011, \apj, 737, 103

\bibitem[{{Schlegel} {et~al.}(1998){Schlegel}, {Finkbeiner}, \& {Davis}}]{1998ApJ...500..525S}
{Schlegel}, D.~J., {Finkbeiner}, D.~P., \& {Davis}, M. 1998, \apj, 500, 525

\bibitem[{{Schr{\"o}der} {et~al.}(2021){Schr{\"o}der}, {van Driel}, \& {Kraan-Korteweg}}]{2021MNRAS.503.5351S}
{Schr{\"o}der}, A.~C., {van Driel}, W., \& {Kraan-Korteweg}, R.~C. 2021, \mnras, 503, 5351

\bibitem[{{Storey-Fisher} {et~al.}(2023){Storey-Fisher}, {Hogg}, {Rix}, {Eilers}, {Fabbian}, {Blanton}, \& {Alonso}}]{2023arXiv230617749S}
{Storey-Fisher}, K., {Hogg}, D.~W., {Rix}, H.-W., {et~al.} 2023, arXiv e-prints, arXiv:2306.17749

\bibitem[{{Valencic} \& {Smith}(2015)}]{2015ApJ...809...66V}
{Valencic}, L.~A. \& {Smith}, R.~K. 2015, \apj, 809, 66

\bibitem[{{Vergely} {et~al.}(1998){Vergely}, {Ferrero}, {Egret}, \& {Koeppen}}]{1998A&A...340..543V}
{Vergely}, J.~L., {Ferrero}, R.~F., {Egret}, D., \& {Koeppen}, J. 1998, \aap, 340, 543

\bibitem[{{Vergely} {et~al.}(2022){Vergely}, {Lallement}, \& {Cox}}]{2022A&A...664A.174V}
{Vergely}, J.~L., {Lallement}, R., \& {Cox}, N.~L.~J. 2022, \aap, 664, A174

\bibitem[{{Vuong} {et~al.}(2003){Vuong}, {Montmerle}, {Grosso}, {Feigelson}, {Verstraete}, \& {Ozawa}}]{2003A&A...408..581V}
{Vuong}, M.~H., {Montmerle}, T., {Grosso}, N., {et~al.} 2003, \aap, 408, 581

\bibitem[{{Watson}(2011)}]{2011A&A...533A..16W}
{Watson}, D. 2011, \aap, 533, A16

\bibitem[{{Willingale} {et~al.}(2013){Willingale}, {Starling}, {Beardmore}, {Tanvir}, \& {O'Brien}}]{2013MNRAS.431..394W}
{Willingale}, R., {Starling}, R.~L.~C., {Beardmore}, A.~P., {Tanvir}, N.~R., \& {O'Brien}, P.~T. 2013, \mnras, 431, 394

\bibitem[{{Wilms} {et~al.}(2000){Wilms}, {Allen}, \& {McCray}}]{2000ApJ...542..914W}
{Wilms}, J., {Allen}, A., \& {McCray}, R. 2000, \apj, 542, 914

\bibitem[{{Yao} {et~al.}(2017){Yao}, {Manchester}, \& {Wang}}]{2017ApJ...835...29Y}
{Yao}, J.~M., {Manchester}, R.~N., \& {Wang}, N. 2017, \apj, 835, 29

\bibitem[{{Yu} {et~al.}(2023){Yu}, {Khanna}, {Themessl}, {Hekker}, {Dr{\'e}au}, {Gizon}, \& {Bi}}]{2023ApJS..264...41Y}
{Yu}, J., {Khanna}, S., {Themessl}, N., {et~al.} 2023, \apjs, 264, 41

\bibitem[{{Zhang} \& {Kainulainen}(2022)}]{2022MNRAS.517.5180Z}
{Zhang}, M. \& {Kainulainen}, J. 2022, \mnras, 517, 5180

\bibitem[{{Zhang} {et~al.}(2023){Zhang}, {Green}, \& {Rix}}]{2023MNRAS.524.1855Z}
{Zhang}, X., {Green}, G.~M., \& {Rix}, H.-W. 2023, \mnras, 524, 1855

\bibitem[{{Zhu} {et~al.}(2017){Zhu}, {Tian}, {Li}, \& {Zhang}}]{2017MNRAS.471.3494Z}
{Zhu}, H., {Tian}, W., {Li}, A., \& {Zhang}, M. 2017, \mnras, 471, 3494

\end{thebibliography}

%
%

%
\end{document}